\documentclass[12pt]{article}  
\newtheorem{lemma}{Lemma}[section]
\newtheorem{theorem}[lemma]{Theorem}
\newtheorem{proposition}[lemma]{Proposition}
\newtheorem{corollary}[lemma]{Corollary}
\newtheorem{remark}[lemma]{Remark}

\newtheorem{definition}[lemma]{Definition}

\def\sq{\hbox {\rlap{$\sqcap$}$\sqcup$}}
\def\sq{\hbox {\rlap{$\sqcap$}$\sqcup$}}
\def\R{ {\rm R \kern -.31cm I \kern .15cm}}
\def\C{ {\rm C \kern -.15cm \vrule width.5pt \kern .12cm}}
\def\Z{ {\rm Z \kern -.27cm \angle \kern .02cm}}
\def\N{ {\rm N \kern -.26cm \vrule width.4pt \kern .10cm}}
\def\1{{\rm 1\mskip-4.5mu l} }
\def\lsim{\raise0.3ex\hbox{$<$\kern-0.75em\raise-1.1ex\hbox{$\sim$}}}
\def\gsim{\raise0.3ex\hbox{$>$\kern-0.75em\raise-1.1ex\hbox{$\sim$}}}
\def\noi{\noindent}

\def\beq{\begin{equation}}   \def\eeq{\end{equation}}
\def\bea{\begin{eqnarray}}  \def\eea{\end{eqnarray}}
\def\nn{\nonumber}
\def\noi{\noindent}
\newcommand{\QED}{\mbox{}\hfill \raisebox{-2pt}{\rule{5.6pt}{8pt}\rule{4pt}{0pt}} 
          \medskip\par}   
\newcommand\mysection{\setcounter{equation}{0}\section}
\renewcommand{\theequation}{\thesection.\arabic{equation}}
\newcounter{hran} \renewcommand{\thehran}{\thesection.\arabic{hran}}

\def\bmini{\setcounter{hran}{\value{equation}}
  \refstepcounter{hran}\setcounter{equation}{0}
  \renewcommand{\theequation}{\thehran\alph{equation}}\begin{eqnarray}}

\def\bminiG#1{\setcounter{hran}{\value{equation}}
\refstepcounter{hran}\setcounter{equation}{-1}
\renewcommand{\theequation}{\thehran\alph{equation}}
\refstepcounter{equation}\label{#1}\begin{eqnarray}}

\def\emini{\end{eqnarray}\relax\setcounter{equation}{\value{hran}}\renewcommand{\theequation}{\thesection.\arabic{equation}}}

\begin{document} 
\pagestyle{empty}
\centerline{\Large\bf Rigorous semiclassical results for the} 
 \vskip 3 truemm \centerline{\Large\bf magnetic response of an
electron gas} \vskip 0.5 truecm

\centerline{\bf Monique Combescure}
\centerline{Laboratoire de Physique Th\'eorique\footnote{Unit\'e Mixte de
Recherche - CNRS - UMR 8627}}  \centerline{Universit\'e de Paris XI, B\^atiment
210, F-91405 ORSAY Cedex, France}
\centerline{Monique.Combescure@th.u-psud.fr}
\vskip 5 truemm \centerline{\bf and}
\vskip 3 truemm
\centerline{\bf Didier Robert}
\centerline{D\'epartement de Math\'ematiques, CNRS UMR 6629}
\centerline{Universit\'e de Nantes, 2 rue de la Houssini\`ere, F-44322 NANTES Cedex
03, France} 
\centerline{Didier.Robert@math.univ-nantes.fr}
\vskip 1 truecm
\begin{abstract}
Consider a free electron gas in a confining potential and a magnetic field in
arbitrary dimensions. If this gas is in thermal equilibrium with a reservoir at
temperature $T >0$, one can study its orbital magnetic response (omitting the spin).
One defines a conveniently ``smeared out'' magnetization $M$, and the corresponding
magnetic susceptibility $\chi$, which will be analyzed from a semiclassical point of view,
namely when $\hbar$ (the Planck constant) is small compared to classical actions
characterizing the system. Then various regimes of temperature $T$ are studied where
$M$ and $\chi$ can be obtained in the form of suitable asymptotic $\hbar$-expansions.
In particular when $T$ is of the order of $\hbar$, oscillations ``\`a la de
Haas-van Alphen'' appear, that can be linked to the classical periodic orbits of the
electronic motion.
\end{abstract}
\vskip 1.5 truecm
\noi LPT Orsay 00-01 \par
\noi August 2000 \par
\newpage
\pagestyle{myheadings}
\baselineskip 18pt
\mysection{Introduction}
\hspace*{\parindent} The magnetic response theory for a free electron gas is an
old problem con\-si\-de\-red by Landau \cite{la}, Fock \cite{foc} and Peierls \cite{pe}.
The revival of interest in physics arose from the advances of recent experiments
that made possible measurements of the magnetic response on small 2-dimensional
electronic devices. These devices are so ``pure'' that the classical as well as
quantum motion inside them can be considered as ``ballistic'', i.e. is uniquely
determined by the confining potential. (Taking into account impurities would
consist in adding the random potential created by random point scatterers inside
the material). The bi-dimensional structure of such electron-gas is realized
through semi-conductor heterostructures whose size, and shape can be controlled
experimentally, together with the number $N$ of confined electrons. The system
being in contact with a reservoir at temperature $T$, and submitted to a
ma\-gne\-tic field $B$ perpendicular to the surface, the magnetic response can be
measured~: say, the magnetization $M$ or magnetic susceptibility $\chi$ as a
function of the thermodynamic parameters $T$, $N$, $B$. \cite{ruj}. These experiments
manifest the sensitivity of the magnetic response to the integrable versus
non-integrable character of the classical dynamics of \underbar{one} electron
in the system. Numerical experiments on two-dimensional magnetic billards have
confirmed this observation, and suggested that the quantum magnetic response is
an experimentally accessible criterion for distinguishing classically integrable
versus chaotic dynamics \cite{lrpw}. A number of theoretical studies have analyzed
the magnetic response from a ``semi-classical'' point of view, namely as
properties manifesting themselves in the limit when $\hbar$ (the Planck
constant) is small compared to classical actions characterizing the system (say
$h \ll a^2eB/c$ where $a$ is a typical size of the system, $e$ the charge of the
electron, $c$ the velocity of light and $B$ the magnetic field size) \cite{bu}. In
these studies, the Coulomb interactions between the electrons in the system are
neglected, so that the system is a ``free electron gas'' to which the usual
thermodynamic formalism is applied. \par
The thermodynamic functions in the
grand-canonical ensemble can be expressed through the density of states of the
quantum Hamiltonian for one electron in the system. This quantum density of
states, in the semi-classical limit, splits into a mean part and a strongly
oscillating one, according to the well known semi-classical trace formula.
This formula is known in mathematics as Poisson formula (Colin de Verdi\`ere \cite{cdv} , 
 Duistermaat-Guillemin \cite{dg}) 
 and in physics as the Gutzwiller trace formula  in the
chaotic case \cite{gu}, or the Berry-Tabor trace formula in the integrable
case \cite{bt}. This splitting provides a similar splitting in the magnetic response, which
allows to understand the oscillations ``\`a la de Haas-van Alphen'' of the
magnetic susceptibility and their link with the classical periodic orbits of the
electronic motion. \par

The aim of the present paper is to reconsider these questions from a
mathematical point of view, in the following two directions (for non-interacting
electron gases in arbitrary dimension, and not necessarily homogeneous magnetic
fields) 

\begin{description}
\item{\quad  -} examine the regimes of temperature in which the magnetic response
can be obtained semi-classically in the form of asymptotic $\hbar$ expansions
 \item{\quad  -} investigate a ``mesoscopic'' regime of low temperatures where the
periodic orbits of the classical one-electron dynamics manifest themselves as
highly oscillating contributions, to the magnetic response. 
\end{description}  

In a recent work, Fournais \cite{fou} studies the semi-classics of the quantum
current for a non-interacting gas of electrons in dimension $n$ and temperature
$T$, confined in a potential $V$ and subject to a suitable magnetic field $B$.
For fixed non-zero temperature $T$, he obtains a complete asymptotic expansion of
the quantum current in small $\hbar$, and for zero temperature, he obtains the
dominant contribution plus an error term under suitable assumptions. J. Butler
\cite{bu} has recently reexamined this last case using a ``semi-classical trace
formula'' by Petkov and Popov \cite{pp}. \par

We recall that in all these studies, the spin of the electron is omitted so that
only the \underbar{orbital} magnetic response is considered. 

The  content of our
paper is the following~:
\begin{itemize}
\item In section 2 we consider the case when the temperature $T$ is large compared to the Planck constant
 $\hbar$. We  prove asymptotic expansion in $\hbar$ for the thermodynamical potential and we recover
 the Landau diamagnetic formula  for 2-dimensional free electron gas.\\
\item In section 3 we consider the case where  the temperature $T$ is of the same order as  $\hbar$. Then
 we prove that the magnetization  splits into two terms~: an average part with a regular asymptotic expansion in $\hbar$
 plus an oscillating part in $\hbar$ which is the contribution of the periodic orbits of the classical motion.\\
\item In section 4 we come back to the regime $T$ larger than $\hbar$ and prove that the contribution of
 non zero periods of  the classical motion is exponentially small in $\hbar$.
\end{itemize}

\mysection{The Landau magnetism}
\hspace*{\parindent} We shall first give the notations and assumptions that will
hold all along this paper. \par

Given $\beta > 0$, we set~: 

\beq F_{\beta}(x) = - {1 \over \beta} \ {\rm Log} \left ( 1 + e^{-\beta
x} \right )\label{2.1}
\eeq

\beq
f_{\beta}(x) = F'_{\beta}(x) = \left ( 1 + e^{\beta x} \right )^{-1}
\label{2.2}
\eeq
$f_\beta$  is related with the Fermi-Dirac distribution. 

\noi These functions are meromorphic, with poles (or cuts for $F_{\beta}$) at~:

$$x = {2k + 1 \over \beta} \ i\pi \qquad k \in \Z$$

\noi Let $\kappa \in {I \hskip - 1 truemm R}$ be a real  parameter (coupling constant with
 a magnetic field). We consider a family
of Hamiltonians with magnetic fields given by~:

\beq
H_\kappa (q, p) = {1 \over 2} \left ( p - \kappa a(q) \right )^2 + V(q)
\label{2.3}
\eeq

\noi where $V$~: ${I\hskip - 1 truemm R}^n \mapsto {I\hskip - 1 truemm R}$ 
and $a$~: ${I\hskip - 1 truemm R}^n \mapsto {I\hskip - 1 truemm R}^n$ 
are ${\cal C}^{\infty}$ functions satisfying the following properties.

$$\forall q\in\R^n,\;\; V(q) \geq 1 \ , \ |\partial_q^{\alpha} V(q) | \leq C_{\alpha} V(q) \leqno({\rm H.1})$$

$$\forall q\in\R^n,\;\; |\partial_q^{\alpha} a(q) | \leq C_{\alpha} \ V(q)^{1/2} \leqno({\rm H.2})$$

$$\forall q\in\R^n,\;\; V(q) \geq c_0 \left ( 1 + |q|^2 \right )^{s/2} \quad \hbox{some $s$, $c_0 > 0$}
\leqno({\rm H.3})$$

\noi (confinement assumption). \par

Let now $\widehat{H}_\kappa$ be the Weyl quantization of $H_\kappa$. The previous
assumptions ensure that  $\widehat{H}_\kappa$ is self-adjoint and its
spectrum $\sigma(\widehat{H}_\kappa)  \subset [\varepsilon, \infty ) $ is pure point 
 for $|\kappa | \leq \kappa_0$ where $\varepsilon >0$ (\cite{ro}). 
Let us call $(E_{j})_{j \in {I \hskip - 1 truemm N}}$ and
$(\varphi_j )_{j \in {I\hskip - 1 truemm N}}$ the set of corresponding
eigenvalues and eigenstates.  \par

In the grand-canonical ensemble, the thermodynamic potential $\Omega$ is given
by~: 

\beq
\label{2.4}
\Omega (\beta , \mu , \kappa) = \sum_{j\in\N}F_\beta(E_j - \mu) = {\rm Tr}\left \{ F_{\beta} \left (\widehat{H}_\kappa - \mu \right )
\right \} \eeq

\noi where $\mu > 0$ is the chemical potential,  $\beta = 1/k_B T$,  $\ k_B$ being
the Boltzmann cons\-tant, and $T>0$ the temperature. ($\kappa$ will be the size of the
magnetic field).  Furthermore, the mean-number of particles in the grand-canonical
ensemble is given by~:

\beq
N(\beta , \mu , \kappa) = {\rm Tr} \left ( f_{\beta} \left ( \widehat{H}_\kappa - \mu \right )
\right ) \eeq

\noi Using the functional calculus \cite{ro}, it is not difficult to see that
$F_{\beta}(\widehat{H} - \mu )$ and $f_{\beta}(\widehat{H} - \mu )$ are
trace-class   and that the function~: $\kappa \mapsto \Omega
(\beta , \mu , \kappa)$ is ${\cal C}^{\infty}$ for $|\kappa| \leq \kappa_0$. \par\noindent
We shall  denote $\partial_\kappa = \frac{\partial}{\partial \kappa}$. 

\begin{proposition}\label{reg} The function~: $\kappa \mapsto \Omega
(\beta , \mu , \kappa)$ is ${\cal C}^{\infty}$  on $\R$. 
 In particular we have
\beq
\partial_\kappa \Omega  = {\rm Tr} \left[f_{\beta}
\left (\widehat{H}_\kappa - \mu \right )\partial_\kappa \widehat{H}_\kappa \right]
\eeq
\end{proposition}

\noindent
Now we have the following definitions of magnetization $M$  and magnetic
susceptibility $\chi$~: 

\bea
\label{mag}
M = \partial_\kappa \Omega  & = & {\rm Tr} \left[ f_{\beta}
\left (\widehat{H}_\kappa - \mu \right ) \partial_\kappa \widehat{H}_\kappa  \right]\\
\label{susc}
\chi & =  & \partial_\kappa M 
\eea
 \noi \underbar{\bf Proof of proposition}\ref{reg}  
Since $\sigma (\widehat{H}_\kappa) \subset [\varepsilon_0, +\infty )$, we can
draw a  suitable curve $\Lambda$ in the complex energy plane, around $\sigma
(\widehat{H}_\kappa)$, with all branching points of $F_{\beta}(z-\mu)$ left
outside. So using Cauchy formula we have 
\beq
F_{\beta}\left ( \widehat{H}_\kappa - \mu \right ) = {1 \over 2i\pi} \int_{\Lambda} dz
\ F_{\beta} (z - \mu ) \left ( \widehat{H}_\kappa - z \right )^{-1} 
\eeq
Using Lebesgue convergence theorem  and cyclicity of the trace, we get
\beq
\partial_\kappa \Omega = 
-{1 \over 2i\pi}{\rm Tr}\int_{\Lambda} dz
\ \left[F_{\beta} (z - \mu ) \left ( \widehat{H}_\kappa - z \right)^{-2}\partial_\kappa\widehat{H}_\kappa\right]
\eeq
Integration by parts give
\beq
\partial_\kappa \Omega = 
{1 \over 2i\pi}{\rm Tr}\int_{\Lambda} dz
f_{\beta} (z - \mu ) \left ( \widehat{H}_\kappa - z \right )^{-1}\partial_\kappa\widehat{H}_\kappa 
\eeq
This procedure can be easily iterated to prove  that $\Omega$ is $C^\infty$-smooth in $\kappa$. Moreover
in the semiclassical regime we can prove that  the asymptotics for  derivatives in $\kappa$  of $\Omega$ can be computed
 using the following commutators formulas for derivatives of the resolvent. Starting from the
 well known identity~:
\beq
[\hat{A}, (\hat{H}-z)^{-1}] = (\hat{H}-z)^{-1}[\hat{H}, \hat{A}](\hat{H}-z)^{-1}, 
\eeq
we get 
\bea
\partial_\kappa(\hat{H}-z)^{-1} = \partial_\kappa\hat{H}(\hat{H}-z)^{-2} - [\partial_\kappa\hat{H},
\hat{H}](\hat{H}-z)^{-3}\nonumber\\
 - [\hat{H}, [\hat{H}, \partial_\kappa\hat{H}]](\hat{H}-z)^{-4} +
 (\hat{H}-z)^{-1}[\hat{H}, [\hat{H}, [\hat{H}, \partial_\kappa\hat{H}]]](\hat{H}-z)^{-4}.
\eea
Each commutator gives one $\hbar$ and we can  compute in the same way higher derivatives in $\kappa$. So using
 Cauchy formula we can compute asymptotics in $\hbar$ of derivatives in $\kappa$ of $\Omega$.
\QED

\begin{theorem}  For any $\varepsilon > 0$ and $\kappa_0 > 0$,   $\Omega$ admits an asymptotic expansion in
$\hbar$, uniform in $\kappa$ for $|\kappa| \leq \kappa_0$, and for
 $\beta \leq \hbar^{\varepsilon- 2/3}$. More explicitly, for any $N \in {I\hskip - 1 truemm N}$ we have ~:

\beq
\label{2.8}
\Omega = h^{-n} \sum_{j=0}^N \sum_{k \leq {3j \over 2}} {(-1)^{k+1} \over k
!} \hbar^j \ \Omega_{jk} + O\left ( \hbar^{N+1-n} \ \beta^{{3N \over 2} + k(n)}
\right ) \eeq

\noi {\it with} $$\Omega_{jk} = \int_{{I \hskip - 1 truemm R}^{2n}} dq \ dp \
d_{jk}(q,p) \ F_{\beta}^{(k)} ( H_\kappa - \mu )$$

\noi {\it $d_{jk}$ being a suitable linear combination of derivatives of $H_\kappa$
with respect to $q$, $p$ and $k(n)$ a constant depending only on the dimension $n$ ($k(n) \leq 2n+1$). In
particular~:} 

\beq
\Omega_{00} = \int_{{I \hskip - 1 truemm R}^{2n}} dq\ dp\ F_{\beta} \left (
H_\kappa(q,p) - \mu \right ) \eeq

\beq
{1 \over 2} \Omega_{22} - {1 \over 6} \Omega_{23} = - {\beta \over 48 \pi^2}
\int_{{I \hskip - 1 truemm R}^{2n}} dq \ dp {\kappa^2\parallel B(q)\parallel^2 -
\sum\limits_{jk} \partial^2_{jk}V \over \cosh^2 \left [ {\beta \over 2} \left ( H_\kappa (q, p) - \mu
\right ) \right ]}  \eeq
 where $B_{jk}$ is the magnetic field

\beq
\parallel B \parallel^2 = \sum_{j < k} B_{jk}^2, \qquad B_{jk} = {\partial a_j
\over \partial q_k} - {\partial a_k \over \partial q_j},\; \partial^2_{jk}V = \frac{\partial^2V}{\partial q_j\partial q_k}.
 \eeq
 and we have chosen the gauge so that $\partial a/\partial q$ is
symmetric. Moreover, the asymptotic expansion can be derived term by term with respect
to $\kappa$ and yields an asymptotic expansion of the magnetization and the magnetic
susceptibility.
\end{theorem}

\noi \underbar{\bf Proof}. We start with the  following
Cauchy formula as in  the proof of proposition (\ref{reg})
\beq
F_{\beta} \left ( \widehat{H}_\kappa - \mu \right ) = {1 \over 2i\pi} \int_{\Lambda} dz
\ F_{\beta} (z - \mu ) \left ( \widehat{H}_\kappa - z \right )^{-1} \eeq

\noi Proceeding as in \cite{ro}, good enough  semi-classical approximations of
$(\widehat{H}_\kappa - z)^{-1}$ (for $z \in \Lambda$) are obtained for any integer $N$ of the following
form~: 
\beq\label{param}
\left ( \widehat{H}_\kappa - z \right )^{-1} = \sum^N_{j=0} \hbar^j \widehat{b_j(z)} -
h^{N+1} \left ( \widehat{H}_\kappa - z \right )^{-1} \ \widehat{R}_N(z) \eeq

\noi where $\widehat{A}$ denotes Weyl quantization, and $b_j(z)$ for $j \geq 2$ are
obtained, from $b_0(z) = (H_\kappa (q, p) - z)^{-1}$ by the formula

\beq
b_j (z) = \sum_{2 \leq \ell  \leq \left [{3j \over 2} \right ]} d_{j\ell} \ b_0^{\ell +1} (z) 
\eeq

\noi $d_{j\ell }$ being a symbol constructed through partial derivatives of $H_\kappa(q,p)$ 
and that can be computed explicitly. Furthermore,
due to the particular form (\ref{2.3}) of $H_\kappa$, $b_j \equiv 0$ for odd $j$'s, and $R_N$ obeys~: 

\beq\label{rem}
|R_N(z)| \leq C_N \left | {z \over Im \ z} \right |^{{3N \over 2} + k(n)}, 
\eeq
 $k(n)$  depending  only on the dimension $n$. (for details see \cite{daro}).\par

Now inserting (\ref{param}) into the Cauchy formula  and using

\beq
f^{(m)}(\lambda ) = {(-1)^m m! \over 2 i \pi } \int_{\Lambda} dz \ f(z) \ (z -
\lambda )^{-m-1}
\eeq

\noi together with $(h = 2 \pi \hbar )$~:

\beq
{\rm Tr} \ \widehat{A} = h^{-n} \int_{{I \hskip - 1 truemm R}^{2n}} dp \ dq \ A(q,
p)
\eeq

\noi we get the result. Let us make explicit the calculus. We
have to find $b_2(z)$ such that 

\beq
\left ( \widehat{H}_\kappa - z \right ) \left ( \widehat{b_0(z)} + \hbar^2
\widehat{b_2(z)} \right ) = 1 + O(\hbar^3) 
\eeq

\noi which, according to the rule for the symbol of the product of two operators,
yields

\beq
\label{2.15}
b_2(z) = d_{22} \ b_0^3(z) + d_{23} \ b_0^4(z)
\eeq

\noi with $b_0(z) = (H_\kappa - z)^{-1}$, and

\beq
\left \{ \begin{array}{l} d_{22} = - \displaystyle{{1 \over 4}} \sum\limits_{|\alpha
| + |\alpha '| = 2} \left ( \partial_p^{\alpha} \partial_q^{\alpha '} H_\kappa \right )
\left ( \partial_p^{\alpha '} \partial_q^{\alpha} H_\kappa \right ) {(-1)^{|\alpha '|}
\over \alpha ! \alpha '!} \\ \\ d_{23} = \displaystyle{{1 \over 2}}
\sum\limits_{|\alpha | + |\alpha '|=2} \left ( \partial_p^{\alpha}
\partial_q^{\alpha '} H_\kappa \right ) \left ( \partial_p^{\alpha '} H_\kappa \right )
\left ( \partial_q^{\alpha} H_\kappa \right ) \displaystyle{{(-1)^{|\alpha '|} \over
\alpha ! \alpha ' !}} \end{array} \right . \eeq

\noi $\alpha$ being a multi-index $\alpha = (\alpha_1, \cdots \alpha_n) \in {I\hskip - 1
truemm N}^n$, we denote as usually by $\partial_p^{\alpha}$ the multiple derivative
${\partial^{\alpha_s} \over \partial p_1} {\partial^{\alpha_2} \over \partial
p_2} \cdots {\partial^{\alpha_n} \over \partial p_n}$, by $|\alpha |$ the sum
$\sum\limits_{j=1}^n \alpha_j$, and by $\alpha !$ the product $\prod\limits_{j=1}^n
\alpha_j !$. \par

Now the calculi proceed as in \cite{hr} ~:

\begin{eqnarray*}
\Omega_{23} &=& {1 \over 4} \int_{{I \hskip - 1 truemm R}^{2n}} dq\ dq \
F_{\beta}^{(3)} (H_{\kappa} - \mu) \left ( \sum_{j,k=1}^n \left ( \partial_{q_j} H_\kappa \right )
\left ( \partial^2_{p_jp_k}H_\kappa \right ) \left ( \partial_{q_k} H_\kappa \right ) \right . \\
&&\left . - 2 \left ( \partial_{q_j}H_\kappa \right ) \left ( \partial_{p_k}H_\kappa \right )
\left ( \partial^2_{p_j q_k}H_\kappa \right ) + \left ( \partial p_j H_\kappa \right )
\left ( \partial_{p_k}H_\kappa \right ) \left ( \partial^2_{q_j q_k} H_\kappa \right ) \right )
\end{eqnarray*}

\noi and  integrating by parts (over $q_j$, or $p_k$) we get

\beq
\label{2.17}
\Omega_{23} = 2 \int_{{I \hskip - 1 truemm R}^{2n}} dq\ dp \ F_{\beta}^{(2)}(H_\kappa -
\mu ) d_{22} (q, p) = 2 \Omega_{22}
 \eeq

\noi and thus~:

\beq
{1 \over 2} \Omega_{22} - {1 \over 6} \Omega_{23} = {1 \over 6} \Omega_{22}
\eeq

\noi We now make $d_{22}$ explicit~:

\bea
d_{22} &=& {1 \over 4} \sum_{j,k} \left ( \partial_{q_jp_k}^2 H_\kappa \right )
\left ( \partial^2_{p_jq_k} H_\kappa \right ) - \left ( \partial^2_{q_jq_k}H_\kappa \right )
\left ( \partial^2_{p_j p_k} H_\kappa \right ) \nn \\
&=& {1 \over 4} \left \{ \kappa^2\parallel B \parallel^2 + \sum_{j\ell} \left ( \kappa\left (
p_j - \kappa a_j(q)\right ) \cdot \partial^2_{j\ell}a(q) - \partial_{j_\ell}V(q) \right ) \right \}  \eea

\noi Clearly the term $(p_j - \kappa a_j(q))\partial^2_{jk}a(q)$ does not contribute to
$\Omega_{22}$ using the change of variable $p \to p - \kappa a(q)$, and the oddness of
the integrand with respect to $p$ variable. Thus we are left with

$$\Omega_{22} = {1 \over 4} \int_{{I \hskip - 1 truemm R}^{2n}} dq\ dp\
F_{\beta}^{(2)} (H_0 - \mu ) \left ( \kappa^2 \parallel B(q) \parallel^2 - \sum_{jk}
\partial^2_{jk}V(q) \right )$$

\noi Now the uniformity of the asymptotic expansion in $\hbar$ with respect to
$\beta \leq \hbar^{\varepsilon - {2 \over 3}}$ (for any  $\varepsilon > 0$) comes
from the fact that $F_{\beta}^{(k)}(x) = \beta^{k-1} F_1^{(k)}(\beta x)$, ($k
\geq 1$) so that $\hbar^j \beta^k \leq \hbar^{j\varepsilon}$ for $k \leq {3j
\over 2}$. Furthermore, the error term in (\ref{2.8}) follows from (2.21).
\par
For the magnetization $M$, we start with formula (\ref{mag})
and we use the semi-classical expansion
 for the resolvant to get 
\bea\label{appM}
M = \frac{1}{2i\pi}{\rm Tr}\int_\Lambda dz f_\beta(z-\mu)
\left(\sum_{0\leq j\leq N}\hbar^j\hat{b_j}(z)\partial_\kappa\widehat{H_\kappa}\right) \\ \noindent
-\frac{\hbar^{N+1}}{2i\pi}{\rm Tr}\int_\Lambda dz f_\beta(z-\mu)(\widehat{H_\kappa} - z)^{-1}\widehat{R_N}(z)\partial_\kappa 
\widehat{H_\kappa}
\eea
Then using  integration by parts we can prove that the first term in (\ref{appM}) is
\beq
\hbar^{-n}\sum_{0\leq j\leq N}\sum_{k\leq 3j/2} \frac{(-1)^{k+1}}{k!}\hbar^j\partial_\kappa \Omega_{j,k},
\eeq
and the second term in (\ref{appM}), using estimate (\ref{rem}), is
\beq
O\left(\hbar^{N+1-n}\beta^{3N/2 + k(n)}\right),
\eeq
uniformly in $\kappa$ for $\vert \kappa\vert \leq \kappa_0$. \\
The same method can be used to prove semiclassical asymptotics for the susceptibility $\chi$ and  also
 for higher order derivatives in $\kappa$ of $\Omega$.
\QED
Now let us denote by $\sum_{\mu}^{\kappa}$ the energy surface at energy $\mu$.
\beq
\sum\nolimits_{\mu}^\kappa = \left \{ (q, p) \in {I \hskip - 1 truemm R}^{2n} : H_\kappa(q,p) = \mu
\right \}
 \eeq
 
\noi and by $d\sigma_{\mu}^\kappa$ the Liouville measure on $\sum_{\mu}^\kappa$~:

\beq
\label{2.21}
d\sigma_{\mu}^\kappa(q,p) = \frac{d \sum_{\mu}^\kappa}{|\nabla H_\kappa|}
\eeq

\noi defined for any non-critical $\mu$ (i.e. $\nabla H_\kappa (q, p) \not= 0$ on
$\sum_{\mu}^\kappa$) ($d\sum_{\mu}^\kappa$ Lebesgue Euclidean  measure). Then we have 
\begin{corollary}[Landau diamagnetism] Let $\chi$ be defined by (\ref{susc}), and $\mu$
be non-critical for $H_0$. Then for $n=2$  and $\kappa = 0$ we have

\beq
\lim_{ \hbar \to 0, \beta \to \infty,  \beta \leq \hbar^{\varepsilon - 2/3}} \chi = - {1 \over 24 \pi^2}
\int_{\sum_{\mu}^0} \parallel B(q)\parallel^2 d\sigma_{\mu}^0
\eeq

\noi which is nothing but Landau's result of the diamagnetism for a 2-dimensional
free electron gas.
\end{corollary}

\mysection{A ``trace formula'' for the magnetization } 
\hspace*{\parindent} For a temperature regime $\beta
\geq \hbar^{-{2 \over 3} + \varepsilon}$ ($\varepsilon > 0$), in order to get
information on the semi-classical limit of magnetization $M$, we shall use
the Fourier inversion formula instead of Cauchy formula. First of all let us remark that $f_\beta$ is not in the
 Schwartz space ${\cal S}(\R)$ so it is more convenient to take its derivative,  which is
 in ${\cal S}(\R)$. We have explicitly 
$$
f_\beta^\prime(x) = \frac{-\beta}{4\cosh^2(\beta x/2)}
$$
So we have the following   Fourier transform formula
\beq\label{fourier}
f_\beta^\prime(x) = -\frac{1}{2\pi}\int_\R dt\frac{\pi t/\beta}{\sinh(\pi t/\beta)}{\rm e}^{itx}
\eeq
So we can write 
\beq
f'_{\beta}(\widehat{H} - \mu ) = h^{-1} \int_{- \infty}^{+ \infty} dt \
e^{it(\widehat{H} - \mu)/\hbar} \ \frac{\pi t/\sigma}{\sinh \pi t/\sigma}
 \eeq

\noi where

\beq
\sigma = \beta {\hbar}
\eeq
The parameter $\sigma$,  which has the dimension of a time,  will be important in what follows. It
 plays a role in the Kubo-Martin-Schwinger condition (see\cite{co}).\\ \noindent
\noi Because we cannot compute the semiclassical evolution for infinite time, we shall consider a ``smeared out'' magnetization 
 defined as follows. 
\noi Fix $\tau_0$ and $\tau$~: $0 < \tau_0 < \tau$~; given a ${\cal C}^{\infty}$ even
function $\rho$ such that ~:

\bea\label{cut}
&&\rho (t) \equiv 1 \qquad {\rm if} \quad |t| \leq 1 \qquad  \nn \\
&&\rho (t) \equiv 0 \qquad {\rm if} |t| \geq 2\;\;  {\rm and}\;  \int_\R \rho(t) dt = 1 
 \eea
Let  us define for $\tau >0$, 
\beq
\rho_{\tau} (t) = \rho (t/\tau )
\eeq
and
\beq
M_{\tau} = {\rm Tr} \left \{ \left ( f_{\sigma} * \widetilde{\rho}_{\tau} \right )
\left (\frac{\widehat{H}_\kappa - \mu}{ \hbar} \right )\partial_\kappa \widehat{H}_\kappa\right\}
\eeq

\noi where $\widetilde{g}$ denotes the inverse Fourier transform of $g$. Clearly
$M_{\tau} \to M$ when $\tau \to \infty$. However we shall not be able to let $\tau
\to \infty$ in this paper (see discussion at the end of this section), and we will
only obtain results for finite $\tau$. \par

Consider a cut-off function $\theta \in {\cal C}_0^{\infty} ({I \hskip - 1 truemm
R})$ with supp $\theta \subset [- \delta , \delta ]$, and $\theta \equiv 1$ on
$[- {\delta \over 2}, {\delta \over 2}]$. It is a priori arbitrary but in the
sequel, we will take $\delta$ so small that if $\mu$ is non-critical for $H_\kappa$,
any $\lambda \in \ ]\mu - 3 \delta , \mu + 3 \delta [$ will remain so. \par

Let us write the following decomposition

\beq
M_{\tau} = M_{\tau , \theta } + M_{\tau , 1 - \theta}
\eeq 

\noi where

\beq
\left \{ 
\begin{array}{l} 
M_{\tau, \theta} = {\rm Tr} \left \{ \left ( f_{\sigma} *
\widetilde{\rho}_{\tau} \right ) \left ( \frac{\widehat{H}_\kappa - \mu}
{\hbar}\right ) \theta \left ( \widehat{H}_\kappa - \mu \right )\partial_\kappa
\widehat{H}_\kappa  \right \} \\ 
\\ 
M_{\tau, 1 - \theta} = {\rm Tr }\left
\{ \left ( f_{\sigma} * \widetilde{\rho}_{\tau} \right ) \left (
\frac{\widehat{H}_\kappa - \mu }{\hbar} \right ) (1 - \theta ) \left (
\widehat{H}_\kappa - \mu \right )\partial_\kappa \widehat{H}_\kappa \right\} 
\end{array} \right . 
\eeq
 We now prove

\begin{lemma}\label{L1} Let us assume (H1-3) for $\widehat{H}_\kappa$, and let $\sigma > \sigma_0 > 0$. 
Then $M_{\tau, 1 - \theta}$ has a
complete asymptotic expansion in $\hbar$.
\end{lemma}

\noi \underbar{\bf Proof}. $(1 - \theta )(x)$ is supported by the union of
$(- \infty , - \delta ]$ and  $[\delta , + \infty )$, 
 which yields two contributions to $M_{\tau, 1 - \theta }$ that we call $M_{\tau, 1 -
\theta }^{\pm}$. \par

Since $f_{\sigma} * \widetilde{\rho}_{\tau} = - \int_x^{\infty} (f'_{\sigma} *
\widetilde{\rho}_{\tau})(y) dy$ is the primitive vanishing at $+ \infty$ of a function in the
Schwartz class ${\cal S}(\R)$, we have for every $N$,  uniformly for
$\sigma > \sigma_0$ and $\hbar\in]0, 1]$, 
$$
\vert{\rm Tr} \left \{ (f_{\sigma} * \widetilde{\rho}_{\tau} ) \left ( {\widehat{H}_\kappa - \mu
\over \hbar} \right ) (1 - \theta )^+ \left ( \widehat{H}_\kappa - \mu \right ) 
\partial_\kappa\widehat{H}_\kappa \right \}\vert \leq
C_N \ \hbar^N 
$$ 

We now consider $M_{\tau , 1 - \theta}^-$. \par

 Since $\int_{-\infty}^{+ \infty} (f'_{\sigma} * \widetilde{\rho}_{\tau}) (y) dy =
(\widetilde{f'_{\gamma}} \cdot \rho_{\tau})(0) = 1$, we clearly have, uniformly
for any $\lambda \in sp (\widehat{H}_\kappa - \mu) \cap (- \infty , - \delta ]$, and any $\sigma >
\sigma_0$~:

\beq\label{1}
f_{\sigma} * \widetilde{\rho}_{\tau} \left ( {\lambda \over \hbar} \right ) = 1 + O(\hbar^N)
\eeq

\noi and, since $\widehat{H}_\kappa$ is semi-bounded from below, the contribution of 1 in
(\ref{1}) gives 
${\rm Tr} \left \{ (1 - \theta )^- (\widehat{H}_\kappa - \mu) \partial \widehat{H}_\kappa \right \}$ 
and obviously has a complete
$\hbar$ expansion by the functional calculus (in fact it is of the form 
${\rm Tr }\left \{ \theta_1 (\widehat{H}_\kappa - \mu)\partial_\kappa
\widehat{H}_\kappa  \right \}$ 
for some $\theta_1 \in {\cal C}_0^{\infty}({I
\hskip - 1 truemm R})$.) $\sq$ \\

The next step is to decompose $\rho_{\tau}$ in order to isolate the neighborhood of $t = 0$ of
the rest~:

\beq
\label{3.9}
\rho_{\tau} = \rho_{\tau_0} \ \rho_{\tau} + \left ( 1 - \rho_{\tau_0} \right ) \rho_{\tau}
\equiv \rho_{\tau_0} + \rho_{1, \tau}
 \eeq 

\noi since $\rho_{\tau_0} \rho_{\tau} = \rho_{\tau_0}$ if $\tau > 2 \tau_0$. \par

This yields $\widetilde{\rho}_{\tau} = \widetilde{\rho}_{\tau_0} + \widetilde{\rho}_{1, \tau}$
and correspondingly~:

\beq\label{M}
M_{\tau , \theta} = M_0 + M_{osc}
\eeq

\beq\label{3.12ref}
\left \{ 
\begin{array}{l} 
M_0 = {\rm Tr} \left \{ \left ( f_{\sigma} *
\widetilde{\rho}_{\tau_0} \right ) \left ( \displaystyle{{\widehat{H}_\kappa - \mu
\over \hbar}} \right ) \theta \left ( \widehat{H}_\kappa - \mu \right ) \partial_\kappa
\widehat{H}_\kappa  \right \} \\ 
\\ 
M_{osc} = {\rm Tr} \left
\{ \left ( f_{\sigma} * \widetilde{\rho}_{1,\tau} \right ) \left (
\displaystyle{{\widehat{H}_\kappa - \mu \over \hbar}} \right ) \theta  \left (
\widehat{H}_\kappa - \mu \right ) \partial_\kappa \widehat{H}_\kappa  \right
\} \end{array} \right .
\eeq

\noi Finally $M_{\tau}$ is decomposed into

\beq
M_{\tau} = \overline{M} + M_{osc}
\eeq

\noi where $M_{osc}$ is given by (\ref{M}), and

\beq\label{MM}
\overline{M} = M_0 + M_{\tau, 1 - \theta}
\eeq

\noi We prove~: \\

\begin{lemma}\label{L2}  Assume also  $\mu$ is non-critical for $H_\kappa$. Then for
  $\tau_0$  small enough  the
classical flow induced by Hamiltonian $H_\kappa$ on $\sum_{\mu}^\kappa$ 
 has no periodic point of period $\leq {\tau_0
\over 2}$  and  $M_0$ admits an asymptotic
expansion in $\hbar$, uniform for $\sigma = \beta {\hbar} \geq \sigma_0 > 0$.
\end{lemma}

\noi \underbar{\bf Proof.} 
We have
\bea\label{M0}
M_0 & = & \int_{+ \infty}^{\mu} d \lambda \ {\rm Tr} \left \{ \left ( f'_{\sigma} *
\widetilde{\rho}_{\tau_0} \right ) \left ( {\widehat{H}_\kappa - \lambda \over \hbar} \right ) \theta
\left ( \widehat{H}_\kappa - \mu \right ) \partial_\kappa \widehat{H}_\kappa  \right\} \nn \\
& = & - \int_{\mu}^{\mu + 3 \delta} d \lambda \ { \rm Tr } \left\{ \left ( f'_{\sigma} *\widetilde{\rho}_{\tau_0} \right )
 \left ({\widehat{H}_\kappa - \lambda} \over \hbar \right ) \theta
\left ( \widehat{H}_\kappa - \mu \right ) \partial_\kappa \widehat{H}_\kappa \right \} +
O(h^{\infty}) \nn \\ 
\eea

\noi using the fact that $f'_{\sigma} * \widetilde{\rho}_{\tau_0}$ is in the space ${\cal S}(\R)$
 and the support property of the cut-off function $\theta$.
Now, if $\mu$ is non-critical for $\widehat{H}_\kappa$, it will remain true for any $\lambda \in
[\mu , \mu + 3 \delta ]$ for small enough $\delta$. \par

We can rewrite the Trace inside the integral in (\ref{M0}), using inverse Fourier
transform,

\beq
M_0 = {1 \over h} \int_{- \infty}^{+ \infty} dt \ {\pi t/\sigma \over \sinh
 \ \pi t/\sigma} \rho \left (
{t \over \tau_0} \right ) {\rm Tr} \left \{ e^{-it (\widehat{H} - \lambda )/\hbar} \ \theta
\left ( \widehat{H} - \mu \right ) \partial_\kappa \widehat{H} \right \}\eeq

\noi (omitting the index $\kappa$ of $\widehat{H}_\kappa$ for simplicity). Now, using either WKB method
(see for instance \cite{ro}), or the coherent state decomposition \cite{cr}, a complete
asymptotic expansion in $\hbar$ of $M_0$  can be obtained, of the form~:

$$ M_0 = \hbar^{-n+1} \left ( C_0(\lambda ) + \hbar C_1 (\lambda ) + \cdots + h^kC_k(\lambda) + \cdots \right ) \qquad {\rm mod} \
O(h^{\infty})$$

\noi which can be further integrated with respect to $\lambda$ on the interval $[\mu , \mu +
3 \delta ]$, yielding the result. $\sq$ \\

\begin{remark} Above Lemmas  therefore imply that $\overline{M}$ 
 has a complete asymptotic expansion in $\hbar$. It is, so to say, the analog of
the (complete $\hbar$-expansion of) the ``mean density of states'' in the Gutzwiller trace
formula. The other term $M_{osc}$  will be the sum of highly oscillating
terms, also in complete analogy with the oscillatory part of Gutzwiller trace formula.
Before showing this now, let us remark that the dominant $\hbar$ contribution to $\overline{M}$
has not yet been shown to reduce to the well known ``Landau diamagnetism''. This will be
postponed to the end of this section.
\end{remark}

\begin{proposition}\label{Mosc}  Assume (H1-3) together with 

\begin{description}
\item{(H.4)} $\mu$ is non-critical for $H_\kappa$ ($|\kappa | < \kappa_0$)
\item{(H.5)} On $\sum_{\mu}^\kappa$, the set $(\Gamma_{\mu})_{\tau}$ of classical periodic orbits
denoted $\gamma$ with period smaller than $\tau$ is such that the corresponding Poincar\'e maps
$P_\gamma$ do not have eigenvalue 1.
 \end{description}   
\noi Then for any $\sigma_1 > 0$ and for $\kappa_0 >0 $ small enough,  
we have the following uniform asymptotics for $\beta\hbar=\sigma\in ]0,\sigma_1]$
 and $\vert \kappa\vert \leq \kappa_0$,
\beq\label{fmosc}
M_{osc} = \sum_{\gamma \in (\Gamma_{\mu})_{\tau}} e^{i(S_\gamma/\hbar + \nu_\gamma {\pi \over 2})} 
\left \{
\frac{\rho_{1, \tau}(T_\gamma)}{|\det (1 - P_\gamma)|^{1/2}} \ \frac{im_\gamma/2 \sigma }{\sinh (\pi T_\gamma/\sigma)}
 + \sum_{k \geq 1}
d_\gamma^{(k)} \hbar^k \right \} + O(\hbar^{\infty})
\eeq
 where $m_\gamma = \int_0^{T_\gamma^*} dt \partial_\kappa \ H_\kappa (q_t, p_t)$. \par

$T_\gamma^*$ is the primitive period of orbit $\gamma$, \par
$S_\gamma$ (resp. $\nu_\gamma$) is the classical action (resp. Maslov index) of orbit $\gamma$. \par
$d_\gamma^{(k)}$ are constants depending on orbit $\gamma$, on the function $\rho_{1, \tau}$, and on
$\gamma$. \\
Moreover the different orbits $\gamma$ can be choosen such that they depend smoothly on the parameter $\kappa$
 and the asymptotic expansion holds uniformly in $\kappa$ for $\vert\kappa\vert$ small enough.
\end{proposition} 

\noi \underbar{\bf Proof}. Since $\rho_{1, \tau}$ is supported away from zero, we can rewrite
$M_{osc}$ (defined in (\ref{3.12ref})) as the non-singular integral~: 

\beq
M_{osc} = {1 \over h}\int_{- \infty}^{+ \infty} {dt \over t} \rho_{1, \tau} (t) \frac{\pi t/\sigma}
{\sinh \ \pi t / \sigma} {\rm Tr} \left \{ \theta \left ( \widehat{H_\kappa} - \mu \right )
e^{-it(\widehat{H_\kappa} - \mu )/\hbar} \ \partial_\kappa \widehat{H_\kappa}  \right \}  
\eeq

\noi Now the method that we have developed in \cite{crr} applies, and yields the desired result.
$\sq$ 

\begin{remark}
At zero magnetic field ($\kappa = 0$)  the term $m_\gamma$ is computed  as follows. 
 We have
$$
m_\gamma = \oint_\gamma \partial_\kappa H_\kappa(q_t, p_t)dt = -\oint_\gamma adq = - \Phi_\gamma
$$
where $\Phi_\gamma$is the flux of the magnetic field through the closed curve $\gamma$.
\end{remark}

We now come back to $\overline{M}$, and prove that the dominant contribution of the
$\hbar$-expansion is indeed the well-known diamagnetic Landau term.

\begin{proposition} Assume (H1-4). Then for any $\sigma_1>0$ and for $\kappa_0>0$,
   uniformly for $\beta\hbar=\sigma\in ]0,\sigma_1[$
we have, mod $O(h^{\infty})$, 

\beq\label{landau}
\overline{M} = - \kappa \frac{\hbar^{2-n}}{24 \pi^2} \int_{\sum_{\mu}^0} d\sigma^0_{\mu} \parallel B(q)
\parallel^2 \ + \sum_{k \geq 3 - n} c_k (\mu , 0, \sigma, T) \hbar^k + O(\hbar^{ \infty}).
\eeq
\end{proposition}

\noi \underbar{\bf Proof}. Whereas the existence of a complete asymptotic expansion for
$\overline{M}$ results immediately from lemmas (\ref{L1}) and (\ref{L2}), the explicit calculus of the dominant
contribution in (\ref{landau}) is not an immediate consequence. It will be done through the
functional calculus. In the previous section we have shown that the coefficients of the
asymptotic $\hbar$-expansion are regular functions of $\kappa$, and can be differentiated with
respect to $\kappa$. Thus, according to (\ref{MM}), we shall obtain the dominant contribution to
$\overline{M}$ from two different contributions $\Omega_0 (\lambda ) |_{\lambda= \mu}$ and
$\Omega_{\tau, 1 - \theta} (\lambda)|_{\lambda = \mu}$ by differentiating with respect to $\kappa$~:

\beq
\left \{ \begin{array}{l} \Omega_0 (\lambda ) = {\rm Tr} \left [ \left ( F_{\beta} *
\widetilde{\rho}_{\tau_0}\right ) \left ( \displaystyle{{\widehat{H}_\kappa - \lambda \over \hbar}}
\right ) \theta \left ( \widehat{H}_\kappa - \mu \right ) \right ] \\ \\ \Omega_{0, 1 - \theta}
(\lambda ) = {\rm Tr}\left [ \left ( F_{\beta} * \widetilde{\rho}_{\tau}\right ) \left (
\displaystyle{{\widehat{H}_\kappa - \lambda \over \hbar}} \right ) ( 1 - \theta ) \left (
\widehat{H}_\kappa - \mu \right ) \right ]\end{array}\right .  \eeq

\noi Let us consider the contribution of  second derivatives in $\lambda$  of $\Omega_0(\lambda )$~:

\beq\label{comp}
\Omega''_0 (\lambda ) := G(\lambda ) = h^{-n} c''_0(\lambda ) + h^{-n+2} c''_2(\lambda ) +
O(h^{-n+3}) \eeq

\noi and we shall identify $c_0(\lambda )$ and $c_2(\lambda )$ by the following trick (inspired
from ref. \cite{ro} prop. V.8)~: Take $\varphi \in {\cal C}_0^{\infty} (] \mu - 3 \delta , \mu +
3 \delta [)$ and integrate against (\ref{comp})~; we get~:

\beq\label{int}
\int d\lambda \ \varphi (\lambda ) \ G(\lambda ) = h^{-1} \int \widetilde{\rho}_{eff} \left ( {
\lambda \over \hbar} \right ){\rm Tr} \left [ \varphi_{\theta} \left ( \widehat{H} - \lambda \right )
\right ] d\lambda 
\eeq

\noi 
where 
$$
\rho_{eff}(t) = \rho_{\tau_0}(t) \frac{\pi t/\sigma}{ \sinh\left(\pi t/\sigma\right)}, 
$$

$\widetilde{\rho}_{eff} (\lambda)$ is its Fourier transform, and

\beq
\varphi_{\theta} (E) := \varphi (E) \ \theta (E + \lambda - \mu )
\eeq

\noi (\ref{int}) follows from~:
$$
G(\lambda ) = {1 \over h} \int dt \ \rho_{eff} (t) \ {\rm Tr} \left [ e^{-it(\widehat{H_\kappa}- \lambda
)/\hbar} \ \theta \left ( \widehat{H_\kappa} - \mu \right ) \right ]
$$

\noi Now (\ref{int}) is rewritten as

\beq
 \int d\lambda \ \varphi (\lambda ) \ G(\lambda ) = 
{1 \over 2 \pi} \int d\lambda \ \widetilde{\rho}_{eff}(\lambda ) \ {\rm Tr} \left [
\varphi_{\theta}
\left ( \widehat{H_\kappa} - \lambda\hbar \right ) \right ] \eeq

\noi which can be developed through Taylor's formula (since integration variable $\lambda $ is in a
compact interval), as~:

\bea
\int d\lambda \ \varphi (\lambda ) \ G(\lambda ) && = 
{1 \over 2 \pi} \sum_{k = 0}^{\infty} {(-1)^k \hbar^k \over k !} \int d\lambda \
\lambda^{k} \ \widetilde{\rho}_{eff}(\lambda ) {\rm Tr} \left [ \varphi_{\theta}^{(k)}
(\widehat{H_\kappa})\right ] \nn \\
&& = \sum_{k=0}^{\infty} {i^k \hbar^k \over k !} \rho_{eff}^{(k)}(0) \ {\rm Tr} \left [ \varphi_{\theta}^{(k)}
(\widehat{H_\kappa}) \right ]
\eea

\noi Actually the term with $k = 0$ is absent since $\rho '_{eff}(0) = 0$ ($\rho$ is an even
function, and so is $\rho_{eff}$). We calculate the coefficients of $\hbar^{-n}$ and $h^{2-n}$
by the functional calculus, like in section 2~:

\bea \label{int2}
 \int d\lambda\varphi(\lambda)G(\lambda) =  &&h^{-n} \left ( \int \varphi_{\theta} (H(q, p)) dp \ dq - {\hbar^2 \over 12} \int \varphi
''_{\theta} (H(q, p)) \left [ \kappa^2 \parallel B \parallel^2 - \triangle V \right ] dq\ dp \right .\nn \\
&&\left . - {\hbar^2 \over 2} \rho''_{eff}(0) \int \varphi ''_{\theta} \left [ H(q, p) \right
] dq \ dp + O(h^3) \right )  \eea

\noi Clearly the first and third terms in (\ref{int2}) are independent on $\kappa$ (through the change
of variable $p \mapsto p - \kappa a(q)$), and we are left with lower order term

\bea
&&- {h^{2 - n} \over 12 \cdot 4 \pi^2} \int dq \ dp \ \varphi ''_{\theta} (H(q, p)) \left [ \kappa^2
\parallel B \parallel^2 - \sum_{1\leq j,k\leq n}\partial^2_{j,k}V\right ] \nn \\
&&= {-h^{2-n} \over 12 \cdot 4 \pi^2} \int d\lambda \ \varphi_{\theta} (\lambda ) {d^2 \over d
\lambda^2} \int_{\sum_{\lambda}^\kappa} \left ( \kappa^2 \parallel B \parallel^2 - \sum_{1\leq j,k\leq n}\partial^2_{j,k}V \right )
d\sigma_{\lambda}^\kappa (q, p) \eea

\noi where we have used integration by parts~: 

$$
\int_{{I \hskip - 1 truemm R}^{2n}} G(q, p) \ \varphi '' (( H_\kappa(q,p)) dq \ dp = \int d\lambda \
\varphi (\lambda ) {d^2 \over d \lambda^2} \left [ \int_{\sum_{\lambda}^{\kappa}} d\sigma_{\lambda}^{\kappa} \
G(q, p) \right ]
$$

 \noi Therefore since the above calculation holds for an arbitrary test
function $\varphi$, we can identify the functions $c_0$ and $c_2(\lambda )$ appearing in
(\ref{comp}), modulo $\kappa$-independent terms as~:

\beq\label{coef}
\left \{ \begin{array}{l} c_0(\lambda ) = 0 \\ \\ c_2(\lambda ) = - \theta (\lambda )
\displaystyle{{\kappa^2 \over 12 \cdot 4 \pi^2}} \int_{\sum_{\lambda}} \parallel B \parallel^2
d\sigma_{\lambda}(q, p) \end{array}\right . \eeq

\noi We can do the same calculus for $\Omega_{\tau , 1 - \theta }$ instead of $\Omega_0$ by
replacing $\tau_0$ by $\tau$ and $\theta$ by $1 - \theta$. This yields a contribution to the
magnetization which, added to that coming from $c_2(\lambda )$ in (\ref{coef}) gives the
dominant Landau term in (\ref{landau}).

We shall extend now the above results to the magnetic susceptibility $\chi$. The statement is the  following
\begin{theorem}Let us assume H-1 to H-5 and $\sigma = \beta\hbar \in ]0, \sigma_1]$
 where $\sigma_1 > 0$  is fixed. For $\chi_\tau = \chi*\rho_\tau$, $\tau >0$,
 we have the decomposition
\beq
\chi_\tau = \overline{\chi} + \chi_{osc}
\eeq
with
\bea
\overline{\chi} & =  & -  \frac{\hbar^{2-n}}{24 \pi^2} \int_{\sum_{\mu}^{\kappa}} d\sigma^{\kappa}_{\mu} \parallel B(q)
\parallel^2 \ + \sum_{k \geq 3 - n} c_{\chi,k} (\mu , \kappa,\sigma, T) \hbar^k + O(\hbar^{ \infty})\\
\chi_{osc} & = & \sum_{\gamma \in (\Gamma_{\mu})_{\tau}} e^{i(S_\gamma/\hbar + \nu_\gamma {\pi \over 2})} 
\left \{
\frac{\rho_{1, \tau}(T_\gamma)}{|\det (1 - P_\gamma)|^{1/2}} \ \frac{r_\gamma m_\gamma^2/2 \sigma }{\sinh (\pi T_\gamma/\sigma)} + 
\sum_{k \geq 1} d_{\chi, \gamma}^{(k)} \hbar^k \right \} + \nonumber\\  && O(\hbar^{\infty})
\eea
where we have used the notations in proposition (\ref{Mosc}) and $r_\gamma = \frac{T_\gamma}{T_\gamma^*}$, $c_{\chi, \gamma}$, $d_{\chi, \gamma}$
 are smooth coefficients depending on  the  periodic orbit  $\gamma$, on $\sigma$, and on function $\rho$.

\end{theorem}
\noi \underbar{\bf Proof}  We use the same cut-off already introduced for the magnetization $M$.
 So we define  in a natural way
\bea
\overline{\chi}  & = & \partial_\kappa M_{\tau_0, \theta} + \partial_\kappa M_{\tau, 1-\theta}  \\
\chi_{osc} & = & \chi_\tau - \overline{\chi}
\eea
Compute first  the term $\chi_{\tau_0, \theta} = \partial_\kappa M_{\tau_0, \theta}$. From the proof
 of proposition (\ref{Mosc})  we get
\beq
\label{chibar1}
 \chi_{\tau_0, \theta} = 
-{1 \over h}\int_0^\mu d\lambda \int_{- \infty}^{+ \infty} dt \ \frac{\pi t/\sigma}{\sinh(\pi t/\sigma)} \rho \left (
{t \over \tau_0} \right )\partial_\kappa{\rm  Tr} \left \{ {\rm e}^{-\frac{it}{\hbar} (\widehat{H_\kappa} - \lambda )} \ \theta
\left ( \widehat{H_\kappa} - \lambda \right ) \partial_\kappa \ \widehat{H_\kappa} \right \}
\eeq
We compute derivative in the parameter $\kappa$ with the following easy consequence of the Duhamel formula
\bea
\partial_\kappa{\rm  Tr} \left \{ {\rm e}^{-\frac{it}{\hbar} (\widehat{H} - \mu )} \ \theta
\left ( \widehat{H} - \mu \right ) \partial_\kappa \ \widehat{H} \right \}  = 
 {\rm  Tr} \left \{ {\rm e}^{-\frac{it}{\hbar} (\widehat{H} - \lambda )} \ \partial_\kappa[\theta
\left ( \widehat{H} - \lambda \right )\partial_\kappa \ \widehat{H}] \right \}+   \nonumber \\
\frac{1}{i\hbar}{\rm  Tr} \left \{\int_0^tds 
\left({\rm e}^{\frac{is}{\hbar}\widehat{H}_\kappa}\partial_\kappa \widehat{H_\kappa}{\rm e}^{-\frac{is}{\hbar}\widehat{H_\kappa}}\right) 
{\rm e}^{-\frac{it}{\hbar} (\widehat{H_\kappa} -\lambda )}\theta
\left ( \widehat{H} - \mu \right ) \partial_\kappa \ \widehat{H_\kappa} \right \}
\eea
Then due to the support property of $\tau_0$, the only stationary points corresponds to the period $T = 0$  and the leading term in $\hbar$
 is given by the first term. The term $\chi_{\tau, 1-\theta} = \partial_\kappa M_{\tau, 1-\theta}$
 is computed in the same way and  the both term combines to yield the asymptotic expansion of
 $\overline{\chi}$.\\ \noindent
For the term $\chi_{osc}$ we start from a formula like (\ref{chibar1})  replacing the time cut-off $\rho_{\tau_0}$
 by the following
 $\rho_{1, \tau} = \rho_\tau(1 -\rho_{\tau_0})$. Hence applying  the methods of \cite{crr} we can compute
 with the stationary phase theorem the contributions  of the periodic trajectories
 with period $T_\gamma \in (\Gamma_\mu)_\tau$.

\begin{remark} In the so-called ``mesoscopic regime'' examined in this
section (i.e. $T = {\hbar \over \sigma k_B}$ for some fixed $\sigma$ having the dimension of time)
, and in the special case of dimension 2, the dominant semi-classical contribution
$M_L$ to $\overline{M}$ and $M_1$ to $M_{osc}$ are of the same order (apart from highly oscillating
factors). A comparison of the corresponding contributions $\chi_L$ and $\chi_1$ to the
susceptibility is made in the physics literature, measuring a factor of 100 for
$\chi_1/\chi_L$ \cite{ruj}. 
\end{remark}

\begin{remark} Thus the magnetic response is a measurable quantity where the
skeleton of the periodic orbits of the classical motion manifests itself clearly~; we have
investigated this effect rigorously and in great generality. Furthermore the
oscillations in (\ref{fmosc}) are a generalization of the well-known de Haas-van Alphen
oscillations of the magnetic response  which are a result of the classical cyclotronic
orbits demonstrated in dimensions 2 and 3, and which can be recovered from (\ref{fmosc}) in the
limiting case where all classsical orbits are of cyclotronic nature ($V = 0$ or quadratic).
\end{remark}

Now,  we  want to comment about the fact that we have only
been able to give semi-classical expansions for ``smeared out magnetizations'' $M_\tau$
instead of the true one $(\tau = \infty )$. For non-zero temperature $T \not= 0$, the
exponential decrease of $\widetilde{f'_{\beta}}(t)$ when $k \to + \infty$ lets us expect that
the Fourier inversion formula (\ref{fourier}) combined with ``trace formulas'' will be enough to
obtain Proposition (\ref{Mosc})  without the $\widetilde{\rho}_{\tau}$ which cuts off time at $|t| \leq \tau$.
We expect that our method using semi-classical evolution estimates for coherent states \cite{cr}
will allow to prove this for $\sigma = \beta {\hbar} > \sigma_0 > 0$ with suitable assumptions
on the classical flow. This is presently under study. However for $T = 0$, the cut-off
$\widetilde{\rho}_{\tau}$ will be necessary to make the sum over periodic orbits finite and
thus convergent, and we cannot expect to get rid of it.\\ \noindent
For the moment,  using estimates proved in \cite{cr}, we can see that it is sufficient
  to control the periods of the classical flow in the time interval $[\tau, c_0\log(\frac{1}{\hbar}])$.
In  \cite{cr}  we have  proved that the semi-classical propagation of coherent states
 is valid in time interval $[- c_0\log(\frac{1}{\hbar}),  c_0\log(\frac{1}{\hbar})]$  for some $c_0>0$.
 So we can  write down the operator ${\rm e}^{-\frac{it}{\hbar} (\widehat{H} - \lambda )} $
 as a Fourier integral operator with a complex phase for $\vert t\vert \leq c_0\log(\frac{1}{\hbar})$.
So we have to compute two terms,  ($H_\kappa = H$),
\bea
  F_1(\hbar, \sigma) := \int_\R dt{\rm  Tr} \left \{ {\rm e}^{-\frac{it}{\hbar} (\widehat{H} - \mu )} \ \theta
\left ( \widehat{H} - \mu \right )\hat{A} \rho\left(\frac{2t}{c_0\log(\frac{1}{\hbar})}\right)R_\sigma(t) \right\}\\
 F_2(\hbar, \sigma) := \int_R dt{\rm  Tr} \left \{ {\rm e}^{-\frac{it}{\hbar} (\widehat{H} - \mu )} \ \theta
\left ( \widehat{H} - \mu \right )\hat{A}\left[1 - \rho\left(\frac{2t}{c_0\log(\frac{1}{\hbar})}\right)\right]R_\sigma(t) \right\}
\eea
where $\hat{A}$ is some quantum observable and  $R_\sigma(t) = \frac{\pi t/\sigma}{\sinh \pi t/\sigma}$.
The term $F_1$ is difficult to check  and we have nothing to say about it here  except that 
 for each time, it is a Fourier integral with a known complex phase
but  it  is difficult to control the stationary phase argument for large times.\\ \noindent
 The term $F_2$  is easily controlled because it contains 
 the  damping factor $R_\sigma$. More precisely we have

\begin{lemma} There  exists $C > 0$ such that for every  $\hbar \in ]0, 1]$
 and $\sigma > 0$ we have easily~: 
\beq
F_2(\hbar, \sigma) \leq C c_0\log(\frac{1}{\hbar})\hbar^{(\pi/\sigma) c_0}.
\eeq
\end{lemma}
So that  $F_2(\hbar, \sigma)$ is negligible for $\frac{c_0}{\sigma}$ large enough.
\mysection{The regime of temperature $\hbar^{1 - \varepsilon} \leq T \leq \hbar^{{2 \over 3} -
\varepsilon}$} 
\hspace*{\parindent} In section 2 we have shown that the functional calculus applies to the
thermodynamical functions in the grand-canonical ensemble and provided asymptotic expansions in
the semi-classical limit provided $T \geq \hbar^{{2 \over 3} - \varepsilon}$ (some $\varepsilon
> 0$). In section 3 we have investigated a rather different temperature regime (called
``mesoscopic'') where $k_B T = \hbar/\sigma$ (some $\sigma > 0$ but finite) where a splitting of
the magnetic response into a ``mean part'' and an ``oscillating part'' appears in the
semi-classical limit. In order to be complete, the ``in-between regime'' is now considered. \\

\begin{theorem} Assume (H1.4). Then the magnetisation $M = \partial_\kappa\Omega$ 
 has for any temperature $T$ satisfying $\hbar^{1 - \varepsilon}
\leq T \leq
\hbar^{{2 \over 3} - \varepsilon}$ (some $\varepsilon > 0$) a complete asymptotic expansion in
$\hbar$ obtained by taking the derivative in $\kappa$ of the formal expansion in $\hbar$
 for $\Omega$  given (\ref{2.8}).
\end{theorem}

\noi \underbar{\bf Proof}. As in section 3 take $\tau_0 > 0$ so small that, the classical flow
induced by $H_\kappa$ has no periodic point with non-zero period $\in [- 2 \tau_0, 2 \tau_0]$, and
take $\rho_{\tau_0}$ as in section 3. Futhermore let $\theta \in {\cal C}_0^{\infty} 
({I \hskip - 1 truemm R})$ be, as in section 3 ($\theta \equiv
1$ on
$\left [-{\delta \over 2}, {\delta \over 2} \right ]$, and $\equiv 0$ on ${I \hskip - 1 truemm
R}\setminus [- \delta , \delta ]$). We decompose $M$ and

\beq
M = M_{\theta} + M_{1 - \theta }
\eeq 

\noi with

\beq
M_{\theta} = Tr \left \{ f_{\beta} \left ( \widehat{H}_\kappa - \mu \right ) \theta \left (
\widehat{H}_\kappa - \mu \right ) \partial_\kappa \widehat{H}_\kappa \right \}
 \eeq

\noi and similarly for $M_{1 - \theta}$. \par

Furthermore~:

\bea\label{Mta}
M_{\theta} &=& - \int_{\mu}^{\infty} d\lambda \ {\rm Tr} \left \{ f'_{\beta} \left ( \widehat{H}_\kappa -
\lambda \right ) \theta \left ( \widehat{H}_\kappa - \mu \right ) \partial_\kappa \widehat{H}_\kappa \right \} \\
&=& - \int_{\mu}^{\infty} d\lambda {1 \over h} \int_{- \infty}^{+ \infty} dt {\pi t/\sigma
\over \sinh \ \pi t/\sigma} {\rm Tr} \left \{ e^{-it(\widehat{H} - \lambda )} \theta \left ( \widehat{H}
- \mu \right ) \partial_\kappa \widehat{H} \right \} \nn \\
&=& M_{\theta, \rho} + M_{\theta , 1 - \rho}
\eea

\noi where we insert, inside the integral over $t$, the partition of unity \\ \noindent
 $1 = \rho_{\tau_0}(t) + (1 - \rho_{\tau_0})(t)$, which yields, correspondingly a splitting of
$M_{\theta}$ into the two contributions. \QED

\begin{lemma} Assuming (H.1-3), $M_{1 - \theta}$ has a complete asymptotic
expansion in $\hbar$.
\end{lemma}

\noi \underbar{\bf Proof}. We can proceed as in the proof of Lemma (\ref{L1}), by splitting $(1 - \theta
) (x)$ into the sum of two disjoint functions $(1 - \theta )^{\pm}$ supported respectively in
$[\delta , + \infty )$ (for $+$ sign) and $(- \infty , - \delta ]$. Since $f_{\beta}$ is the
primitive vanishing at $+ \infty$ of a function in the Schwartz class of ${\cal C}^{\infty}$
functions of rapid decrease, we have

$$
\vert{\rm Tr} \left \{ f_{\beta} \left ( \widehat{H_\kappa} - \mu \right ) (1 - \theta )^+ \left ( \widehat{H_\kappa}
- \mu \right ) \partial_\kappa \widehat{H_\kappa}  \right \}\vert \leq C_N \ \hbar^N \qquad
\hbox{(for any $N$)}
$$

\noi and
$$
\vert{\rm Tr} \left \{ (1 - f_{\beta}) \left ( \widehat{H} - \mu \right ) (1 - \theta )^- \left (
\widehat{H} - \mu \right ) {\partial \widehat{H} \over \partial \kappa} \right \}\vert \leq C_N \ \hbar^N
\qquad \hbox{( for sany $N$)}
$$

\noi Finally, we know, like in the proof of Lemma (\ref{L1}), that 
${\rm Tr}\left \{(1 - \theta )^-(\widehat{H} - \mu) \partial_\kappa \widehat{H}\right\}$ has a
complete asymptotic
$\hbar$ expansion by the functional calculus. \QED

\begin{lemma} Assuming (H1-3), one has, for $\hbar^{1 - \varepsilon} \leq T
\leq \hbar^{{2 \over 3} - \varepsilon}$
$$M_{\theta , 1 - \rho} =  O \left ( e^{-c_1/\hbar^{\varepsilon}} \right )$$

\noi where $c_1$ is a positive constant only depending on $\tau_0$.
\end{lemma}

\noi \underbar{\bf Proof}. Using (\ref{Mta}), the support property of $\theta$, and the
exponential decrease of $f'_{\beta}$, it is easy to show that~:

$$M_{\theta} = - \int_{\mu}^{\mu + 2 \delta} d \lambda \ {\rm Tr} \left \{ f'_{\beta}\left (
\widehat{H_\kappa} - \lambda \right ) \theta \left ( \widehat{H_\kappa} - \mu \right ) \partial \widehat{H_\kappa}
 \right \} + O\left ( e^{-\delta / \sqrt{\hbar}} \right )$$

\noi Therefore

$$
M_{\theta, 1 - \rho} = - \int_{\mu}^{\mu + 2 \delta } d\lambda h^{-1} \int_{-\infty}^{+
\infty} dt \left ( 1 - \rho_{\tau_0} (t) \right ) {\pi t/ \sigma \over \sinh \ \pi t/ \sigma}
$$
$$
{\rm Tr} \left \{ {\rm e}^{-it (\widehat{H} - \lambda )/\hbar} \theta \left ( \widehat{H} - \mu \right )
\partial_\kappa \widehat{H_\kappa} \right \} + O \left ( {\rm e}^{-\delta /\sqrt{\hbar}} \right)
$$ 

\noi and since, in the considered temperature regime 
$$\left | {\pi t/ \sigma \over \sinh (\pi t/\sigma)} \right | \leq C \
{\rm e}^{-c_0|t|/\hbar^{\varepsilon}}
$$

\noi we have, using the support property of $ 1 - \rho_{\tau_0}$~:

$$\left | M_{\theta , 1 - \rho} \right | \leq C \ e^{-c_1/\hbar^{\varepsilon}}$$

\noi $c_1$ being a positive constant depending on $\tau_0$. \QED 

\begin{lemma} Assuming (H.1-4), then $M_{\theta ,\rho}$ has a
complete asymptotic expansion in $\hbar$.
\end{lemma}

\noi \underbar{\bf Proof}. As is the previous section, we take $\delta$ so small that, if
$\mu$ is non-critical for $H_\kappa$, then any $\lambda \in [\mu , \mu + 2 \delta]$ is also
non-critical for $H_\kappa$. \par

Now, using the support property of $\rho_{\tau_0}$, and either WKB method, or decomposition
over coherent states, a complete asymptotic expansion can be obtained for   

$$
h^{-1} \int_{- \infty}^{+ \infty} dt \ \rho_{\tau_0}(t) {\pi t / \sigma \over \sinh \ \pi t/
\sigma} {\rm Tr} \left \{ e^{-it(\widehat{H}_{\kappa} - \lambda ) / \hbar} \ \theta \left ( \widehat{H}_{\kappa} - \mu
\right )  \widehat{H}_{\kappa} \right \}
$$

\noi for any $\lambda \in [\mu , \mu + 2 \delta]$. Integrating with respect to $\lambda$ in this
interval yields the result. \QED

\end{document}